\documentclass[prl,twocolumn,preprintnumbers]{revtex4}

\newcommand{\be}{\begin{equation}}
\newcommand{\ee}{\end{equation}}
\newcommand{\bea}{\begin{eqnarray}}
\newcommand{\eea}{\end{eqnarray}}
\newcommand{\bref}[1]{(\ref{#1})}
\newcommand{\LM}{\mathcal{L}_m^{(\mathcal{M})}}
\newcommand{\LS}{\mathcal{L}_m^{(\Sigma)}}
\newcommand{\LGB}{\mathcal{L}_\mathrm{GB}}

\newcommand{\av}[1]{\left< #1 \right>}

\begin{document}

\title{\bf Generalised Israel Junction Conditions for a 
Gauss-Bonnet Brane World}
\author{Stephen C. Davis}
\email{Stephen.Davis@th.u-psud.fr}
\affiliation{LPT, Universit\'e Paris--XI, 
B\^atiment 210, F--91405 Orsay Cedex, France}

\begin{abstract}
In spacetimes of dimension greater than four it is natural to consider
higher order (in $R$) corrections to the Einstein equations. 
In this letter generalized Israel junction conditions for a membrane in
such a theory are derived. This is achieved by generalising the
Gibbons-Hawking boundary term. The junction conditions are applied to
simple brane world models, and are compared to the many contradictory
results in the literature.
\end{abstract}

\preprint{LPT-ORSAY-02-77, {\tt hep-th/0208205}}

\maketitle

It is well known that for a boundary-less spacetime, the Einstein
equations can be derived from the Einstein-Hilbert action. This is no
longer true if the spacetime has a boundary. The problem is resolved by
adding a Gibbons-Hawking boundary term to the action~\cite{GH}. Varying
the action then gives the correct field equations, as well as boundary
conditions at the edge of the spacetime.

A slight variation of this is to consider an infinitely thin $(D-2)$-brane in
a $D$ dimensional spacetime. Since spacetime is split into two, the
brane can then be treated as the boundary of each half of the
spacetime.  Varying the Gibbons-Hawking term then gives the Israel
junction conditions on the membrane~\cite{Israel}.  These junction
conditions have recently received a great deal of attention due to their
use in the study of `brane-worlds' (see for
example~\cite{earlybw,CR,RS,BWC,BWmove}). In a brane-world scenario our
universe is modeled by a 3-brane embedded in a five-dimensional `bulk'
spacetime. In the simplest cases, all the standard model fields are
confined to the brane, while only gravity propagates in the bulk. The
brane is usually taken to be of zero thickness, and so the Israel
junction conditions can be used to relate the bulk dynamics to what we,
on the brane, observe.

The bulk gravitational field equations are usually assumed to be the
five-dimensional Einstein equations. However in spaces of dimension
greater than four it is natural to consider additional higher order
curvature terms~\cite{Lovelock,earlygb,Myers,gbsol}. 

In general relativity, the vacuum field equations are taken to be a
linear combination of the Einstein tensor and the metric. This choice is
motivated by the fact that it is the most general combination of tensors
which is (a) symmetric, (b) depends only on the metric and its first two
derivatives, (c) is divergence free and (d) is linear in second
derivatives of the metric. In fact, in four dimensions, the fourth
condition is superfluous since it is implied by the other
three~\cite{Lovelock}. In five dimensions the second order Lovelock
tensor
\bea
H_{ab} =  R R_{ab} - 2 R_{ac}R^c{}_b
-2 R^{c d}  R_{acbd} + R_a{}^{c d e} R_{b c d e}
&&\nonumber \\ \hspace{0.3in}
{}-\frac{1}{4} g_{ab} \left( R^2 - 4 R_{cd}R^{cd} + R^{cdes} R_{cdes}\right)
&&
\eea
also satisfies the above conditions. Thus the most general choice of
gravitational vacuum field equations in five dimensions is a linear
combination of $g_{ab}$, $G_{ab}$ and $H_{ab}$. In the absence of any
experimental evidence to the contrary, all three terms should be
included.

A further motivation for higher order curvature terms is that they also appear
in the low energy effective field equations arising from most string
theories. Since brane worlds are motivated by string theories, it is
particularly natural to include the extra terms in the five-dimensional
field equations.

Like the standard Einstein equations, these higher order generalisations
can also be derived from an action. The tensor $H_{ab}$ can be
obtained from an action containing the Gauss-Bonnet term
\be
\LGB = R^2 - 4 R_{ab}R^{ab} + R^{abcd} R_{abcd} \ .
\label{LGB}
\ee
Consider the action
\be
S_\mathcal{M} = \frac{1}{2\kappa^2}\int_\mathcal{M} d^Dx\sqrt{-g}  
\left\{ R -2\Lambda +  \alpha \LGB \right\} 
\label{SM}
\ee
for a $D$ dimensional manifold $\mathcal{M}$.
In a string theory context we would have 
$\kappa^{-2} = M_*^{D-2}$ and $\alpha \propto M_*^{-2}$, where $M_*$ 
is the string mass scale.

Let us suppose, as would be the case in a co-dimension one brane world
scenario, that $\mathcal{M}$ is split into two
parts by a hypersurface $\Sigma$, whose two sides will be
denoted $\Sigma_\pm$. Their normals, $n^a$, will be taken to point away
from the surface and {\em into} the adjacent space.

Varying the action~\bref{SM} with respect to the metric gives
\bea
\delta S_\mathcal{M} = \frac{1}{2\kappa^2}\int_\mathcal{M} d^Dx \sqrt{-g} \,
\delta g^{ab} (G_{ab} + \Lambda g_{ab} + 2\alpha H_{ab})
&& \nonumber \\ 
{}-\frac{1}{\kappa^2}\int_{\Sigma_\pm} \! \! d^{D-1}x \sqrt{-h} \, 
n_a\Bigl(g^{a[c}g^{d]b}
+ 2\alpha P^{abcd} \Bigr)\nabla_{\! d} \delta g_{bc} &&
\label{dSM}
\eea
where $h_{ab} = g_{ab} - n_a n_b$ is the induced metric on $\Sigma$, and
\be 
P_{abcd} = R_{abcd} + 2 R_{b[c} g_{d]a} - 2 R_{a[c} g_{d]b} + R g_{a[c}g_{d]b} 
\label{Pdef}
\ee
is the divergence free part of the Riemann tensor.

Expression~\bref{dSM} contains normal derivatives of the metric
variation. As with the Einstein-Hilbert action, we must add a boundary
term in order to cancel them. For an action with a Gauss-Bonnet
term~\bref{SM}, the appropriate term is~\cite{Myers}
\be
S_\Sigma = -\frac{1}{\kappa^2}\int_{\Sigma_\pm} d^{D-1}x \sqrt{-h} \, 
\left(K +2\alpha \{J-2 \widehat G^{ab}K_{ab}\}\right)
\label{SS}
\ee
where $K$ is the trace of the extrinsic curvature, defined by
$K_{ab} = h^c{}_a \nabla_{\! c} n_b$, and $J$ is the trace of
\bea
&&J_{ab} = \frac{1}{3}\bigl(2K K_{ac}K^c{}_b + K_{cd}K^{cd} K_{ab} 
\nonumber \\ && \hspace{0.7in} 
{}- 2K_{ac}K^{cd}K_{db} - K^2 K_{ab} \bigr) \ ,
\eea
Throughout this paper I will denote tensors associated with the
induced metric by a circumflex accent, so
$\widehat G_{ab}$ is the $(D-1)$ dimensional Einstein tensor on $\Sigma$
corresponding to $h_{ab}$.

Varying the action $S_\mathcal{M} + S_\Sigma$ 
now gives an expression which does not contain normal
derivatives of $\delta g_{ab}$.
If we also include a matter contribution to the action 
\be
S_\mathrm{mat} = -\int_\mathcal{M} d^Dx\sqrt{-g}\LM 
- \int_\Sigma d^{D-1}x \sqrt{-h} \LS \ ,
\label{Smat}
\ee
then the variation of the total action 
$S = S_\mathcal{M} + S_\Sigma + S_\mathrm{mat}$ gives
\be
G_{ab} + 2\alpha H_{ab} +\Lambda g_{ab}= \kappa^2 T_{ab}
\ee
in $\mathcal{M}$, and 
\be
2\bigl< K_{ab}-K h_{ab}\bigr> + 4\alpha
\bigl<3 J_{ab} - J h_{ab} +  2\widehat P_{acdb} K^{cd}\bigr> 
= -\kappa^2 S_{ab}
\label{bc}
\ee
on $\Sigma$, with $\av{X} = [X(\Sigma_+)+ X(\Sigma_-)]/2$
denoting the average of a quantity over the two
sides ($\Sigma_\pm$) of the hypersurface. The two energy-momentum tensors
are defined by $T_{ab} = 2\delta \LM/\delta g^{ab} - g_{ab} \LM$ and 
$S_{ab} = 2 \delta \LS/\delta h^{ab} - h_{ab} \LS$.

With the aid of the Gauss-Codazzi equations~(\ref{gc1}--\ref{gc3})
below, we obtain the energy-momentum conservation equation on the hypersurface
\be
D^a S_{ab} = -2\av{n^a(G_{ac} + 2\alpha H_{ac})h^c{}_b} = 
-2\kappa^2\av{n^a T_{ac}h^c{}_b} \ ,
\label{emcon}
\ee
with $D^a$ denoting the covariant derivative corresponding to
$h_{ab}$. This is very similar to the corresponding result in the
standard brane world models~\cite{BWC}.

Many previous papers have tried to derive the junction conditions by
treating the hypersurface $\Sigma$ as a $\delta$-function contribution
to the to $\LM$, as in the original brane cosmology papers~\cite{BWC}. 
In this case there is some ambiguity as to the correct definition of $H_{ab}$
on $\Sigma$, which has led to the suggestion that the junction conditions
must depend on the thickness of the brane~\cite{Nathalie}. This would be
true for a general combination of second order curvature terms, whose
action would contain third (or higher) order derivatives of the
metric. However, this is not true for the Gauss-Bonnet
combination~\bref{LGB} since it has been specifically chosen not to
contain such derivatives.

By allowing $n^c \partial_c g_{ab}$ to be discontinuous at the brane, and
treating $(n^c \partial_c)^2 g_{ab}$ as a $\delta$-function, junction
conditions which are independent of the brane thickness can be
found~\cite{port,nocos,CD,yesdw,yesphi}. However, care must be taken to
regularise the $\delta$-function correctly~\cite{CD,dreg}. This was done
in refs.~\cite{CD,yesdw,yesphi}, and the resulting
junction conditions agree with those in is letter. Refs.~\cite{port,nocos}
do not regularise the $\delta$-function appropriately, and obtain
incorrect results.

We will now use the above results to derive the Friedmann equation for a
cosmological brane world model. As has been shown by Kraus and
Ida~\cite{BWmove} (for the
standard five dimensional Einstein equations), it is possible to obtain a
cosmological generalisation of the Randall-Sundrum~\cite{RS} model by
considering a static bulk spacetime, and allowing the brane to have a
time dependent position in the bulk.

The bulk metric can be written in the form
\be
ds^2 = -h(r) dT^2 + \frac{dr^2}{h(r)} + r^2 \Omega_{ij} dx^i dx^j
\ee
where $\Omega_{ij}$ is the three dimensional metric of space with
constant curvature $k=-1,0,1$. For $\LM=0$, the field equations are
solved by~\cite{gbsol}
\be
h=k+\frac{r^2}{4\alpha}\left(1 - \sqrt{1+\frac{4}{3}\alpha \Lambda 
+ 8\alpha \frac{\mu}{r^4}}\right)
\ee
with $\mu$ being an arbitrary constant. In the $\alpha \to 0$ limit,
$\mu$ is equal to the black hole mass.

We define the position of the brane as $r=a(\tau)$ and $T=t(\tau)$,
which is parameterised by the proper time on the brane $\tau$. The
induced metric is
\be
ds^2 = -d\tau^2 + a(\tau)^2 \Omega_{ij} dx^i dx^j \ ,
\ee
the tangent vector of the brane is $u^a = (\dot T,0,0,0,\dot r)$ and
$n_a = (-\dot r,0,0,0,\dot T)$. Normalisation of $n^a$ implies
\be
-h^2 \dot T^2 + \dot r^2=-h \ .
\label{norm}
\ee
We take the brane matter to be a perfect fluid, so 
$S_{ab} = (\rho+p)u_a u_b +p h_{ab}$. 
The $(uu)$ component of \bref{bc} is then
\be
\left(1 + \frac{8}{3}\alpha H^2 +4\alpha \frac{k}{r^2} \right)
\frac{\bigl< h \dot T \bigr>}{r}
- \frac{4}{3}\alpha\frac{\bigl< h^2 \dot T\bigr>}{r^3}
= -\frac{\rho}{6} \ .
\label{bcuu}
\ee
For simplicity let us assume $Z_2$ symmetry across the brane. Squaring
\bref{bcuu} and simplifying with \bref{norm} gives a cubic equation for
$H^2$. This has the real solution
\be
H^2 = -\frac{k}{a^2} + \frac{c_+ + c_- - 2}{8\alpha}
\label{fried}
\ee
where
\be
c_\pm = \left(\sqrt{
\left(1+\frac{4}{3}\alpha \Lambda + 8\alpha\frac{\mu}{a^4}\right)^{3/2}
+\frac{\alpha \rho^2}{2}} \pm \rho \sqrt{\frac{\alpha}{2}}\right)^{2/3}
\ .
\ee
This Friedmann equation agrees with ref.~\cite{CD}, but not
ref.~\cite{port,nocos}. The only difference between these two conflicting
results is one factor of 3, but surprisingly this gives a substantially
different Friedmann equation. For $\rho = $constant, eq.~\bref{fried}
also agrees with ref.~\cite{yesdw}. Ref.~\cite{altBcos} uses different 
boundary terms~\cite{altBder} to obtain a different Friedmann equation. However
the terms used do not give a consistent action, and so the result is incorrect.

As in the usual brane cosmology~\cite{BWC}, the standard Friedmann
equation can be recovered at late time (large $a$) by splitting $\rho$
into a cosmological constant and a matter part~\cite{CD}.

The boundary terms~\bref{SS} are easily generalised to actions where
other fields couple to the curvature tensors. Variation of the action
\bea
&& \hspace{-0.3in}
S = \frac{1}{2\kappa^2}\int_\mathcal{M} d^Dx\sqrt{-g}  
\left( \Phi(x^\mu) R  -2\Lambda + \alpha \Psi(x^\mu) \LGB \right)
\nonumber \\ && {}
-\frac{1}{\kappa^2}\int_{\Sigma_\pm} d^{D-1}x \sqrt{-h} \, \bigl(\Phi(x^\mu) K 
\nonumber \\ && \hspace{0.5in} 
{}+ 2\alpha  \Psi(x^\mu) \{ J-2 \widehat G^{ab}K_{ab} \} \bigr)
+S_\mathrm{mat}
\label{Sphi}
\eea
produces the field equations
\bea
&&\Phi G_{ab}  - \nabla_{\! a} \nabla_{\! b} \Phi + g_{ab} \nabla^2 \Phi 
+ \Lambda g_{ab}
\nonumber \\ && \hspace{0.3in}
{}+ 2 \alpha \Psi H_{ab} + 4 \alpha P_{eacb} \nabla^e \nabla^c \Psi 
= \kappa^2 T_{ab}
\eea
and the boundary conditions
\bea
&& \hspace{-0.3in}
\left< \Phi (K_{ab}-K h_{ab}) - h_{ab} n^e \partial_e \Phi\right>
\nonumber \\ && 
{}+ 2\alpha\bigl<3\Psi J_{ab} -\Psi J h_{ab} 
+ 2\Psi\widehat P_{acdb} K^{cd} \bigr>
\nonumber \\ && 
{}+ 2\alpha \bigl< \bigl\{2\widehat G_{ab} + 2 K_{ea}K^e{}_b - 2 K K_{ab}  + 
\nonumber \\ && \hspace{0.8in} 
h_{ab} [K^2 - K_{cd}K^{cd}]\bigr\}n^e \partial_e \Psi \bigr>
\nonumber \\ &&
{}+ 8\alpha \bigl< 
\bigl(K_{a[c}h_{b]d} + K_{c[a}h_{d]b} -K h_{a[c}h_{b]d} \bigr) D^c D^d \Psi
\bigr>  
\nonumber \\ && \hspace{0.8in}
= -\frac{1}{2} \kappa^2 S_{ab} \ .
\label{pbc}
\eea
All the problematic boundary terms, like those appearing in \bref{dSM},
cancel out.
In a string theory context $\Phi$ and $\Psi$ would typically be
functions of the dilaton or moduli fields.

If we consider a conformally flat spacetime of the form
\be
ds^2 = - e^{2A(y)} (dT^2 + dx^i dx_i)+ dy^2
\ee
and take $\Phi$, $\Psi$ and $\LS = \lambda/\kappa^2$ to be functions of
$y$ only, the generalised Israel junction conditions reduce to
\be
2\av{3\Phi A' + \Phi'-4\alpha A'^2(A'\Psi+3\Psi')} = -\lambda
\ee
This agrees with other results in the literature~\cite{yesphi}.

\section*{Acknowledgements}

I am grateful to Jihad Mourad and Christos Charmousis for useful and
interesting discussions. This work was supported by EC network
HPRN--CT--2000--00152.

\section*{Appendix}

To prove equation~\bref{bc}, we first use the Gauss-Codazzi equations
\be
R_{pqrs} h^p{}_a h^q{}_b h^r{}_c h^s{}_d =
\widehat R_{abcd} + K_{bc} K_{ad} - K_{ac} K_{bd}
\label{gc1}
\ee
\be
n^a R_{aqrs} h^q{}_b h^r{}_c h^s{}_d = D_d K_{bc} -  D_c K_{bd}
\label{gc2}
\ee
\be
\widehat R_{bd} =
R^a{}_{qcp} h^c{}_a h^q{}_b h^p{}_d + K K_{bd} - K_{bc} K^c{}_d
\label{gc3}
\ee
and contractions of them to expand $n^a P_{abcd}$ in terms of $K_{ab}$,
$n^a$ and quantities associated with the induced metric:
\bea
&& \hspace{-0.3in}
n^a P_{abcd} = 2 D_{[d} K_{c]b} +2D_e K^e{}_{[d}h_{c]b} + 2h_{b[d}D_{c]} K 
\nonumber \\ && \hspace{0.6in}
{}+ 2\widehat G_{b[c}n_{d]} + 2(K_b{}^e - K h_b{}^e) K_{e[c} n_{d]}
\nonumber \\ && \hspace{0.6in}
{}+ (K^2 -K_{ae} K^{ae}) h_{b[c} n_{d]} \ .
\eea
The covariant derivative $D$ is defined by $D_a X_{bc\cdots} = h^m{}_a
h^p{}_b h^q{}_c \cdots \nabla_{\! m} X_{pq\cdots}$.

To find the variation of $S_\Sigma$~\bref{SS} with respect to $g_{ab}$,
we first note that the normalisation of $n_a$ implies 
$\delta n_a = \frac{1}{2} n_a n^c n^d \delta g_{cd}$. Thus, after a
little algebra, and using the definitions of $K_{ab}$ and $D_a$, we obtain
\bea
&& \hspace{-0.3in}
h^c{}_a h^d{}_b\delta K_{cd} =  
n^e\nabla_{\! [e} \delta g_{p]q} h^p{}_{(a} h^q{}_{b)} 
- \frac{1}{2} \delta g^{cd} K_{c(a} h_{b)d}
\nonumber \\ && \hspace{1in}
{}- \frac{1}{2} D_{(a}\left(h^c{}_{b)} \delta g_{ce} n^e\right)
\eea
and, from~\bref{gc3},
\bea
&& \hspace{-0.3in}
h^c{}_a h^d{}_b \delta \widehat R_{cd} =
D_{(a} D^e(h^c{}_{b)} h^d{}_e \delta g_{cd})
\nonumber \\ && 
{}- \frac{1}{2} D^e D_e (h^c{}_a h^d{}_b \delta g_{cd})
- \frac{1}{2} D_a D_b (h^{cd}\delta g_{cd}) \ .
\eea
Now, with the aid of integration by parts and the relation
\bea
&&Y^{abe} \nabla_{\! e} \delta g_{ab} = D_e(Y^{abe} \delta g_{ab})
+ 2n^c \delta g_{cd} Y^{(dp)e} K_{pe} 
\nonumber \\ && \hspace{1in} {}
{}+ \delta g^{ab} D^e Y_{abe} \ ,
\eea
which holds for any tensor $Y^{abe}$ which is orthogonal to the
normal $n^c$, the variation of the total action $S$ can be reduced to
eq.~\bref{bc}. The more general case~\bref{pbc} can be dealt with in the
same way.

The energy conservation equation~\bref{emcon} can be proved by using
$D^a \widehat P_{acdb} =0$ and the Gauss-Codazzi
equations~(\ref{gc1}--\ref{gc3}) to express the divergence of \bref{bc}
in terms of bulk tensors.


\begin{thebibliography}{99}
\bibitem{GH}
G.~W.~Gibbons and S.~W.~Hawking,
Phys.\ Rev.\ D {\bf 15} (1977) 2752.
\bibitem{Israel}
W.~Israel,
Nuovo Cim.\ B {\bf 44} (1966) 1.
\bibitem{earlybw}
K.~Akama,
Lect.\ Notes Phys.\  {\bf 176} (1982) 267 [{\tt hep-th/0001113}]; \\
V.~A.~Rubakov and M.~E.~Shaposhnikov,
Phys.\ Lett.\ B {\bf 125} (1983) 136; \\
M.~Visser,
Phys.\ Lett.\ B {\bf 159} (1985) 22 [{\tt hep-th/9910093}]; \\
E.~J.~Squires,
Phys.\ Lett.\ B {\bf 167} (1986) 286.
\bibitem{CR}
H.~A.~Chamblin and H.~S.~Reall,
Nucl.\ Phys.\ B {\bf 562} (1999) 133 [{\tt hep-th/9903225}].
\bibitem{RS}
L.~Randall and R.~Sundrum,
Phys.\ Rev.\ Lett.\  {\bf 83} (1999) 4690 [{\tt hep-th/9906064}].
\bibitem{BWC}
P.~Binetruy, C.~Deffayet and D.~Langlois,
Nucl.\ Phys.\ B {\bf 565} (2000) 269 [{\tt hep-th/9905012}]; \\
P.~Binetruy, C.~Deffayet, U.~Ellwanger and D.~Langlois,
Phys.\ Lett.\ B {\bf 477} (2000) 285 [{\tt hep-th/9910219}].
\bibitem{BWmove}
P.~Kraus,
JHEP {\bf 9912} (1999) 011 [{\tt hep-th/9910149}]; \\
D.~Ida, 
JHEP {\bf 0009} (2000) 014 [{\tt gr-qc/9912002}].
\bibitem{Lovelock}
D.~Lovelock,
J.\ Math.\ Phys.\ {\bf 12} (1971) 498.
\bibitem{earlygb}
J.~Madore,
Phys.\ Lett.\ A {\bf 110} (1985) 289;
Phys.\ Lett.\ A {\bf 111} (1985) 283; \\
N.~Deruelle and J.~Madore,
Phys.\ Lett.\ A {\bf 114} (1986) 185;
Phys.\ Lett.\ B {\bf 186} (1987) 25; \\
J.~T.~Wheeler,
Nucl.\ Phys.\ B {\bf 268} (1986) 737.
\bibitem{Myers}
R.~C.~Myers,
Phys.\ Rev.\ D {\bf 36} (1987) 392.
\bibitem{gbsol}
D.~G.~Boulware and S.~Deser,
Phys.\ Rev.\ Lett.\  {\bf 55} (1985) 2656; \\
R.~G.~Cai,
Phys.\ Rev.\ D {\bf 65} (2002) 084014 [{\tt hep-th/0109133}].
\bibitem{Nathalie}
N.~Deruelle and T.~Dolezel,
Phys.\ Rev.\ D {\bf 62} (2000) 103502 [{\tt gr-qc/0004021}].
\bibitem{CD}
C.~Charmousis and J.~F.~Dufaux,
to appear in Class.\ Quant.\ Grav.\ [{\tt hep-th/0202107}].
\bibitem{yesdw}
K.~A.~Meissner and M.~Olechowski,
Phys.\ Rev.\ Lett.\  {\bf 86} (2001) 3708
[{\tt hep-th/0009122}].
\bibitem{yesphi}
I.~Low and A.~Zee,
Nucl.\ Phys.\ B {\bf 585} (2000) 395 [{\tt hep-th/0004124}]; \\
N.~E.~Mavromatos and J.~Rizos,
Phys.\ Rev.\ D {\bf 62} (2000) 124004 [{\tt hep-th/0008074}]; \\
I.~P.~Neupane,
JHEP {\bf 0009} (2000) 040 [{\tt hep-th/0008190}]; \\
N.~E.~Mavromatos and J.~Rizos,
{\tt hep-th/0205299}; \\
P.~Binetruy, C.~Charmousis, S.~C.~Davis and J.~F.~Dufaux,
Phys.\ Lett.\ B {\bf 544} (2002) 183 [hep-th/0206089].
\bibitem{port}
C.~Germani and C.~F.~Sopuerta,
Phys.\ Rev.\ Lett.\  {\bf 88} (2002) 231101 [{\tt hep-th/0202060}].
\bibitem{nocos}
J.~E.~Kim, B.~Kyae and H.~M.~Lee,
Nucl.\ Phys.\ B {\bf 582} (2000) 296, 
Erratum: {\em ibid.} {\bf 591} (2000) 587 [{\tt hep-th/0004005}]; \\
B.~Abdesselam and N.~Mohammedi,
Phys.\ Rev.\ D {\bf 65} (2002) 084018 [{\tt hep-th/0110143}].
\bibitem{dreg}
Y.~M.~Cho, I.~P.~Neupane and P.~S.~Wesson,
Nucl.\ Phys.\ B {\bf 621} (2002) 388 [hep-th/0104227].
\bibitem{altBcos}
J.~E.~Lidsey, S.~Nojiri and S.~D.~Odintsov,
JHEP {\bf 0206} (2002) 026 [hep-th/0202198].
\bibitem{altBder}
A.~D.~Barvinsky and S.~N.~Solodukhin,
Nucl.\ Phys.\ B {\bf 479} (1996) 305 [gr-qc/9512047]; \\
M.~Cvetic, S.~Nojiri and S.~D.~Odintsov,
Nucl.\ Phys.\ B {\bf 628} (2002) 295 [hep-th/0112045].
\end{thebibliography}
\end{document}